\documentclass[aps,prl,twocolumn,superscriptaddress]{revtex4-1}
\usepackage{times}
\usepackage{amssymb}
\usepackage{graphicx}
\usepackage{color}
\usepackage{amsmath}
\usepackage{mathtools}
\usepackage{epsfig}
\usepackage{wrapfig}
\usepackage{csquotes}
\usepackage{tikz}
\usetikzlibrary{shapes}
\usepackage{subfigure}
\usepackage{graphics}
\usepackage{graphicx}
\usepackage{color}
\usepackage{amsmath}
\usepackage{graphicx,color}
\usepackage{amsmath}
\usepackage{mathtools}
\usepackage{epsfig}
\usepackage{wrapfig}
\usepackage{csquotes}
\usepackage{cancel} 
\usepackage{stmaryrd}
\usepackage{physics}
\usepackage{soul}
\usepackage{float}

\begin{document}

\title{Rearrangement of 2D aggregates of droplets under compression: \\
signatures of the energy landscape from crystal to glass}
\date{\today}
\author{Jean-Christophe~Ono-dit-Biot}
\affiliation{Department of Physics \& Astronomy, McMaster University, Hamilton, ON, L8S 4L8, Canada}
\author{Pierre~Soulard} 
\affiliation{UMR CNRS Gulliver 7083, ESPCI Paris, PSL Research University, 10 rue Vauquelin, 75005 Paris, France}
\author{Solomon~Barkley}
\affiliation{Department of Physics \& Astronomy, McMaster University, Hamilton, ON, L8S 4L8, Canada}
\author{Eric~R.~Weeks}
\affiliation{Department of Physics, Emory University, Atlanta, GA 30322, USA}
\author{Thomas~Salez} 
\affiliation{Univ. Bordeaux, CNRS, LOMA, UMR 5798, F-33405 Talence, France}
\affiliation{Global Station for Soft Matter, Global Institution for Collaborative Research and Education, Hokkaido University, Sapporo, Japan}
\author{Elie~Rapha\"el} 
\affiliation{UMR CNRS Gulliver 7083, ESPCI Paris, PSL Research University, 10 rue Vauquelin, 75005 Paris, France}
\author{Kari~Dalnoki-Veress} 
\email{dalnoki@mcmaster.ca}
\affiliation{Department of Physics \& Astronomy, McMaster University, Hamilton, ON, L8S 4L8, Canada}
\affiliation{UMR CNRS Gulliver 7083, ESPCI Paris, PSL Research University, 10 rue Vauquelin, 75005 Paris, France}

\begin{abstract}
We study signatures of the energy landscape's evolution through the crystal-to-glass transition by compressing 2D finite aggregates of oil droplets.  Droplets of two distinct sizes are used to compose small aggregates in an aqueous environment.  Aggregates range from perfectly ordered monodisperse \emph{single }crystals to disordered bidisperse glasses.  The aggregates are compressed between two parallel boundaries, with one acting as a force sensor. The compression force provides a signature of the aggregate composition and gives insight into the energy landscape.  In particular, crystals dissipate all the stored energy through single catastrophic fracture events whereas the glassy aggregates break step-by-step. Remarkably, the yielding properties of the 2D aggregates are strongly impacted by even a small amount of disorder.
\end{abstract}

\maketitle

\newcommand{\markercirc}{\raisebox{0.5pt}{\tikz{\node[draw,scale=0.4,circle,fill=black](){};}}}
\newcommand{\markersquare}{\raisebox{0.5pt}{\tikz{\node[draw,scale=0.4,regular polygon, regular polygon sides=4,fill=black](){};}}}

Glassy materials are drastically different from crystals in their properties and cannot simply be described as crystals with defects~\cite{Philips1981}. The intrinsic disorder associated with molecules that do not neatly pack, or polydisperse colloidal spheres, prevents glasses from crystallizing~\cite{Pusey1987,Auer2001}. Intense  effort has been devoted to understanding glasses and the transition from an ordered crystal to a disordered glass. Microscopic properties such as the packing configuration can be accessed experimentally and provide insight into the crystal-to-glass transition~\cite{Yunker2010,Yunker2014,Higler2013,Hanifpour2014}. But,  these studies did not yield any conclusion regarding the difference in mechanical properties between crystals and glasses. To answer this question several numerical studies have been conducted, with a consistent conclusion: adding even a small amount of disorder to a system with crystalline packing results in properties that are similar to amorphous structures~\cite{OHern2003,Goodrich2014,Mari2009,Tong2015,Babu2016,Zhangh2017,charbonneau2018,mizuno2013}. However, conducting an equivalent experimental study is  challenging. A beautiful experiment by Keim {\it et al.} showed that a small amount of disorder in a colloidal poly-crystal results in a shear modulus similar to the one observed with a binary mixture of colloids~\cite{Keim2015}. However, an experimental characterization of the transition from a perfectly ordered single crystal to a disordered glass probed using mechanical properties is still lacking, since experimental systems are often polycrystalline and their properties dominated by grain boundaries.  Here, we experimentally study the yielding properties of 2D finite-size aggregates of droplets that vary in the extent of disorder from a perfect crystal to a glass.

We use an emulsion since individual particles can easily be imaged to obtain both structural and dynamical information~\cite{Weeks2000,Crocker1996,Kose1973}. Colloids and emulsions are proven  model systems for the study of glasses and jamming~\cite{Weeks2000,Jorjadze2011,Hunter2012,Illing2016,Vivek2017}, force chains~\cite{Edwards2003,Desmond2013}, and phase transitions in crystals~\cite{Li2016}. Specifically, we use an emulsion of oil in water (model soft spheres with negligible friction, and an attractive potential)  confined to  a 2D finite-size aggregate. The amount of disorder is tuned by changing the relative fraction of large and small droplets in aggregates with a total of $N_\textrm{tot}=20$ or 23 droplets. We investigate the transition from a perfectly ordered monodisperse crystal~\cite{Pusey1986} to a disordered bidisperse glass~\cite{Zhang2009,Lynch2008,Assoud2009} by systematically adding defects to the crystalline structure. The transition is studied through the force required to globally compress and fracture the 2D aggregates while simultaneously monitoring microscopic reorganization.  The adhesion energy between particles exceeds the thermal energy, thus the particles must be treated as athermal, and the analogue molecular system is that of a glass or a crystal well below the solid-melt transition temperature. The small 2D system of athermal particles provides a unique opportunity to i) prepare perfect single crystals, rather than poly-crystal as is typical, ii) add defects to the single crystals \emph{one-by-one}, and iii) obtain aggregate-scale force response during compression, while iv) \emph{simultaneously}  capturing local structural re-arrangements. With the addition of even a small number of defects, we find: 1) a rapid increase in the number of fracture events upon compression; and 2) that the yield energy is distributed over many small steps in comparison to a single large step for a crystal. These experimental findings provide a signature of the increasingly complex energy landscape as the system transitions from crystal to glass.  An analytical model is developed which supports the experimental data.

\section*{Results and Discussion}
\begin{figure}[h]
 \begin{center}
 \includegraphics[width=.9\columnwidth]{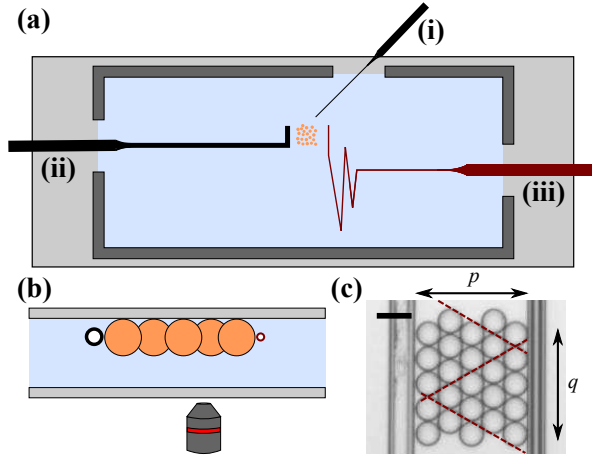}
 \caption{\label{fig1} (a) Schematic top view of the experimental chamber. The typical dimensions of the wall (dark grey) are $55 \times 30 \times 2.5$~mm.  The ``droplet pipette", ``pushing pipette", and ``force-sensing pipette" are labelled as (i), (ii) and (iii) respectively.  (b) Schematic side view (not to scale). The buoyant droplets form a quasi 2D aggregate bounded by the top glass plate. The pushing pipette (black circle on the left) and the force-sensing pipette (red circle on the right) are placed near the average equatorial plane of the droplets so forces are applied horizontally. (c) Optical microscopy image of a typical crystal (scale bar is 50~$\mu$m). Red dashed lines show observed fracture lines for a crystal when compressed.}
 \end{center}
 \end{figure}

The experiment is illustrated in Fig.\ref{fig1}(a) and is described in full detail in the Materials and Methods section. In short, 2D aggregates of buoyant oil droplets are compressed between two thin glass pipettes.  By monitoring the deflection of the ``force sensing pipette'' [see (iii) in Fig.\ref{fig1}(a)] during the compression, the force applied on the aggregate is measured~\cite{Backholm2013, Backholm2019}. The aggregate rearranges under compression by breaking adhesive bonds between droplets. These fracture events can be directly monitored with  optical microscopy and related to the force measurement. We use $p_{\textrm{ini}}$ to refer to the initial number of rows of droplets, defined as parallel to the pipettes as shown in Fig.~\ref{fig1}(c), while $q_{\textrm{ini}}$ refers to the initial number of droplets per row. Under compression the aggregate rearranges to have $p$ rows and $q$ columns, while $N_\textrm{tot}$ remains fixed. Using two ``droplet pipettes'' with different tip radii facilitates the preparation of well controlled bidisperse aggregates~\cite{Barkley2016}. To increase the disorder in an aggregate, large droplets are replaced by small droplets (or {\it vice versa}). 

\subsection*{Effect of Disorder on the Force Curves}
\begin{figure}[]
\begin{center}
\includegraphics[width=0.9\columnwidth,keepaspectratio=true]{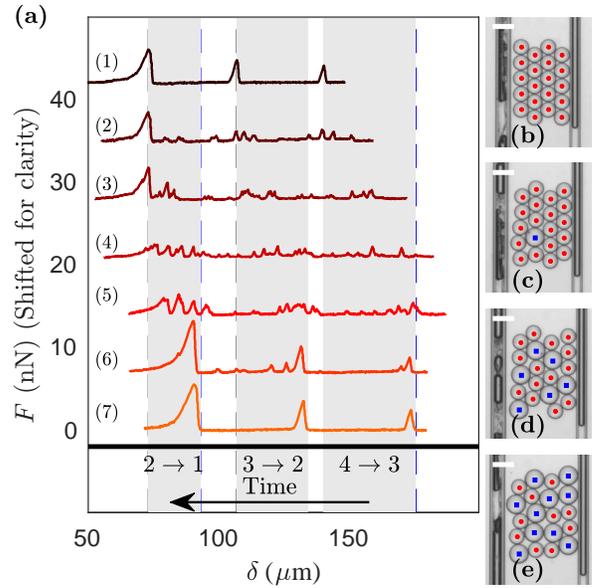}
\caption{\label{fig2} (a) Force measurements, $F$, as a function of the distance between the pipettes, $\delta$, for seven aggregates sharing the same lattice but with different compositions of small ($\mathcal{R}=19.2 \pm 0.3$~$\mu$m) and large ($\mathcal{R}=25.1 \pm 0.3$~$\mu$m) droplets. Here $\delta$ decreases with time as the aggregate is compressed and the aggregate changes from $p=4$ to 1 as indicated at the bottom. Traces 1 to 7 correspond to \{0; 1; 6; 10; 14; 19; 20\} large droplets with $N_\textrm{tot}= 20$. The black dashed lines correspond to the positions $\delta_{\textrm{min}}^p$ of the peak maxima for the crystal made of small droplets, while the blue dashed lines correspond to the positions $\delta_{\textrm{max}}^p$ of the peak onsets (\emph{i.e.} upon compression as $\delta$ decreases) for the crystal made of large droplets. The shaded area highlights the different transitions during the compression. (b-e) Optical microscopy images of the aggregates, before compression, corresponding to traces 1 to 4. Blue squares correspond to large droplets and red circles to small droplets (scale bar is 50~$\mu$m).} 
\end{center}
\end{figure}

In Fig.~\ref{fig2}(a) are shown the force measurements as a function of the distance between the pipettes, $\delta$, for seven different aggregates  with $p_{\textrm{ini}}=4$ and $q_{\textrm{ini}}=5$. The proportion of large and small droplets is varied from aggregate to aggregate. The top trace (1) corresponds to a crystal (\textit{i.e.} a monodisperse aggregate) made of small droplets with radius $\mathcal{R}=r=19.2 \pm 0.3$~$\mu$m [Fig.~\ref{fig2}(b)], and the bottom trace (7) to a crystal of large droplets with radius $\mathcal{R}=R=25.1 \pm 0.3$~$\mu$m.  These traces show three force peaks corresponding to three fracture events: the transition from $p=4$ to $p=3$, which we designate as $4 \rightarrow 3$, followed by $3 \rightarrow 2$, and finally $2 \rightarrow 1$. The peak height is directly linked to the number of bonds broken. Each fracture event corresponds to a local maximum in the force-distance curve of Fig.~\ref{fig2}(a) and to a corresponding inter-basin barrier in the energy landscape. Clearly, for a $p\rightarrow (p-1)$ transition, a crystal made of small droplets will fracture at a smaller spacing between the pipettes (trace 1), compared to a crystal of larger droplets (trace 7). All the bonds are broken in a catastrophic and coordinated manner, in agreement with other studies of crystals under compression~\cite{Gai2016,McDermott2016}. For 2D crystals we find that the fracture patterns consist of equilateral triangles with $(p-1)$ droplets on a triangle's side as shown in Fig.~\ref{fig1}(c). These equilateral triangles arise because they minimize the number of broken bonds  between droplets as $p \rightarrow (p-1)$. After fracture, the triangles slide past each other and reassemble into a new crystal with  $(p-1)$ rows of droplets. By design, the force sensor does not register a friction force during sliding, nor are we sensitive to viscous drag during compression, because slow compression (0.3~$\mu$m/s) ensures that viscous drag forces are negligible.

With the introduction of defects in the structure compression forces are no longer homogeneously distributed within the aggregate (see Fig.~\ref{fig2}(c-e)). Thus, rather than a single catastrophic fracture, additional fracture events occur and extra peaks appear in the force data, as seen in the traces 2 to 6 of Fig.~\ref{fig2}(a). When a single defect is introduced (traces 2 and 6), extra peaks are observed but peaks corresponding to the fracture of the crystalline portion of the aggregate can still be identified (large peaks at the same values of $\delta$). Defects are systematically introduced up to trace 4, which corresponds to the most disordered system that we use to model a glass (equal fraction of large and small droplets). The force-distance curves are strongly impacted by increasing disorder: i) the number of peaks increases; ii) the overall magnitude of the force peaks decreases; and iii) the peaks corresponding to the underlying crystalline structure can no longer be differentiated from the others. 
In order to identify peaks as corresponding to a specific transition from $p$ to $(p-1)$ one can invoke the fact that a restructuring event in a bidisperse aggregate must occur within the compression range set by the \emph{onset} of fracture associated with a crystal of big droplets and the \emph{completion} of fracture in an aggregate of small droplets. Thus, we invoke the following criterion: a peak corresponds to the $p \rightarrow(p-1)$ transition if the peak is found in the compression range $\delta \in [\delta_{\textrm{min}}^p,\delta_{\textrm{max}}^p]$; where $\delta_{\textrm{max}}^p$ is defined by the onset of the force peak upon compression of aggregates only made of large droplets and $\delta_{\textrm{min}}^p$ is defined by the completion of the fracture event in aggregates made of only small droplets  [\emph{i.e.} corresponding to the compression value at the maximum force as detailed in Fig.~\ref{fig2}(a)].

\begin{figure}[t]
\begin{center}
\includegraphics[width=0.9\columnwidth,keepaspectratio=true]{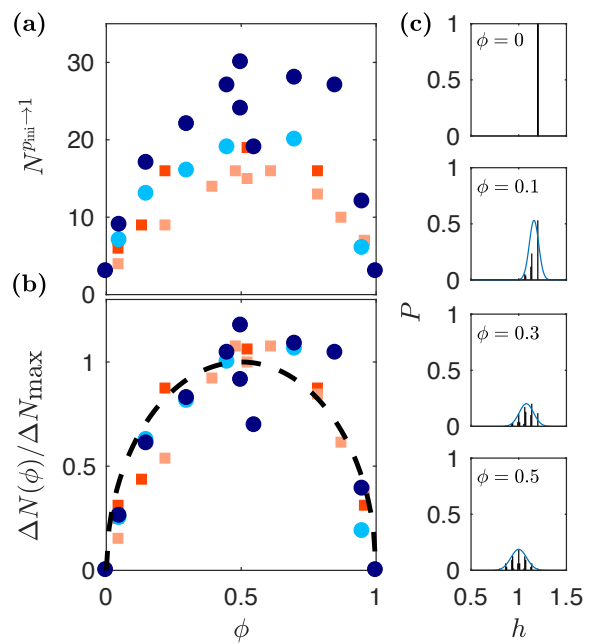}
\caption{\label{fig4} (a) Measured total number of force peaks as a function of defect fraction, for a compression from $p_{\textrm{ini}}$ to $p=1$. (\protect\markercirc) $p_{\textrm{ini}}=4$ with $ q_{\textrm{ini}}=5$ - two data sets (different colours); (\protect\markersquare) $p_{\textrm{ini}}=3$ with the three rows initially made of 8 - 7 - 8 droplets respectively - two data sets (different colours). (b) Evolution of the normalized excess number of peaks compared to a crystal, with the black dashed line corresponding to Eq.~\ref{renorm}. (c)  Theoretical probability distribution of the dimensionless column height $h$, in an aggregate with $q\rightarrow\infty$ and $p=4$, for four defect fractions $\phi = \{0; 0.1; 0.3; 0.5\}$ (see SI). Gaussian curves (blue solid lines) with same standard deviation, $\sigma$, and average, $\mu$, as the discrete distribution are overlayed as a guide to the eyes. Typical radii of large and small droplets are $\approx 22 \ \mu$m and $\approx 18 \ \mu$m (see SI).}
\end{center}
\end{figure}

From Fig.~\ref{fig2} it is evident that fracture properties are strongly dependent on the aggregate composition. In the simplest case, that of compressing a crystal cluster, the droplets deform and the stored elastic energy increases with compression. Eventually the stored elastic energy exceeds the depletion-induced adhesive energy, and a coordinated fracture occurs as discussed above (shown in Figure~\ref{fig1}(c)), such that a minimal number of bonds are broken. We now turn to the more complex bidisperse aggregates. As defects are introduced, the most striking feature is the rapid increase in the number of force peaks [Fig.~\ref{fig2}(a)]. To further quantify this observation, we perform experiments for two different aggregate geometries: i) $p_{\textrm{ini}}=4$ with $ q_{\textrm{ini}}=5$, and ii) $p_{\textrm{ini}}=3$ with the three rows initially made of 8 - 7 - 8 droplets. The composition of the cluster is given by the number fraction of small droplets in  the aggregate, $\phi = N_\textrm{small}/N_\textrm{tot}$, which varies from zero to one. Both $\phi=0$ and $\phi=1$ correspond to crystals while $\phi =0.5$ corresponds to the maximum amount of disorder -- a model glass. The defects are purposely distributed throughout the whole structure to avoid clumps of defects. In Fig.~\ref{fig4}(a), we plot the total (\textit{i.e.} until we reach $p=1$) number $N^{p_\textrm{ini}\rightarrow 1}$ of detectable force peaks as a function of the defect fraction.

\subsection*{Model} We propose a minimal model that rationalizes the experimental observations. A given $(p, q)$ aggregate is approximated by an ensemble of $q$ independent columns, of $p$ rows of droplets each. By allowing this simplification, one can treat each column as a random packing of droplets belonging to two different species which  correspond to the two different radii: small droplets with $\mathcal{R}=r$ and large droplets with $\mathcal{R}=R$. Since droplets are arranged in a nearly hexagonal lattice, each column consists of alternating layers of a single droplet or two droplets side-by-side (see SI for more details).  The probabilities associated with finding a small or large droplet at a specific site are given by the number fractions $\phi$ and $1-\phi$. The total resulting height, $H$, of a given column depends on the specific composition in that column, and takes values ranging from $H_r$ to $H_R$, for columns made of small ($\phi=1$) and large ($\phi=0$) droplets. We define the dimensionless height $h=2H/(H_R+H_r)$. One can compute (see  SI) the associated probability distribution, $P(h)$, plotted in Fig.~\ref{fig4}(c) for various $\phi$ (black bars). Compression of an aggregate then proceeds as follows: First, the tallest columns are compressed and broken, which creates a force peak whose magnitude reflects the abundance of these highest columns in the aggregate. Then, the pushing pipette starts compressing the second highest columns and the process repeats. 

The simple model predicts that the average number of force peaks observed during the $p\rightarrow p-1$ transition of an aggregate can be identified with the average number $N^{p}(\phi,q)$ of different column heights present in the aggregate composed of $q$ columns. For a monodisperse aggregate, there is only one possible column height, and thus $N^{p}(0,q)=N^{p}(1,q)=1$ resulting in one force peak for the $p\rightarrow p-1$ transition. In contrast, as the defect fraction increases, the number of possible different heights and thus the number of force peaks increase. The number of different heights can be calculated numerically according to the scheme described above (see SI). In addition, a simple argument provides an analytical estimate for the average number of force peaks in a sample with a given $\phi$. The increase in the average number of different column heights in comparison to a crystal is expected to be proportional to the standard deviation, $\sigma(\phi,p)\propto\sqrt{\phi(1-\phi)}$, of the height distribution centered at $\mu(\phi,p)$ shown in Fig.~\ref{fig4}(c). This results from the random packings of
the columns described above and gives:
\begin{equation}
N^{p}(\phi,q)-1 =[N^{p}(\phi = 0.5,q) - 1]2\sqrt{\phi(1-\phi)}\ . 
\label{eqpp1}
\end{equation}
Finally, in order to determine all the force peaks encountered on average as the aggregate is compressed, we sum Eq.~(\ref{eqpp1}) over all the transitions starting from a cluster with $p_{\textrm{ini}}$ rows to one row, in order to construct $N^{p_{\textrm{ini}}\rightarrow 1}_{q_{\textrm{ini}}}(\phi)=\sum_{p=2}^{p_{\textrm{ini}}}N^{p}(\phi,q)$, where $q=N_{\textrm{tot}}/p$. Defining the average number of peaks compared to a crystal, $\Delta N(\phi,p_{\textrm{ini}},q_{\textrm{ini}}) = N^{p_{\textrm{ini}}\rightarrow 1}_{q_{\textrm{ini}}}(\phi) - N^{p_{\textrm{ini}}\rightarrow 1}_{q_{\textrm{ini}}}(0)$, we obtain (see SI):
\begin{equation}
\frac{\Delta N}{\Delta N_\textrm{max}}(\phi) = 2\sqrt{(1-\phi)\phi} \ ,
\label{renorm}
\end{equation}
where $\Delta N_{\textrm{max}}=\Delta N(\phi=0.5,p_{\textrm{ini}},q_{\textrm{ini}})$ corresponds to the average maximum excess number of peaks, observed when compressing the most disordered aggregate. The experimental value of $\Delta N_{\textrm{max}}$ is obtained by fitting Eq.~\ref{renorm} to each set of data presented in Fig.~\ref{fig4}(a). Figure~\ref{fig4}(b) shows that this simple model captures well the rapid increase of the number of force peaks as defects are added. The derivative of Eq.~\ref{renorm} at $\phi =0$ is infinite; thus, a small change in the fraction of defects in an aggregate results in a drastic change in the yield properties as observed in experiments. We note that the minor discrepancy between the data and the model reflects experimental error as well as three main departures of the real aggregate from the proposed idealization: i) neighbouring columns are not independent, ii) the real aggregate has a finite number of columns, and iii) some peaks may not be detected. 

\subsection*{Probing the Energy Landscape through the Crystal-to-Glass Transition} The compression experiments can also be used to characterize the yield energy of the aggregate as a function of the defect fraction, which reflects the evolution of the underlying energy landscape through the crystal-to-glass transition. Specifically, the work $W_\textrm{tot}$ exerted (and then fully dissipated in the fluid) in order to generate a $p\rightarrow(p-1)$ rearrangement is obtained by integrating the force-distance curve (Fig.~\ref{fig2}(a)), for the corresponding transition. As explained previously, a $p\rightarrow(p-1)$ transition corresponds to $\delta \in [\delta_{\textrm{min}}^p,\delta_{\textrm{max}}^p]$, so the integration is performed over this interval. Moreover, we only consider the rising (along the compression orientation \emph{i.e.} upon decreasing $\delta$) elastic part $F_{\textrm{s}}$ of the force peaks, as the subsequent decay corresponds to the viscous relaxation of the force-sensing pipette. For this analysis, we focus on the collection of force traces presented in Fig.~\ref{fig2}(a) and in particular the transition $p = 4 \rightarrow p=3$. Within our resolution, the total work $W_\textrm{tot}=\int^{\delta_{\textrm{max}}^p}_{\delta_{\textrm{min}}^p}\textrm{d}\delta'\,F_{\textrm{s}}(\delta') = 2.2 \pm 0.7 \textrm{ fJ}$ is found to be nearly constant for all the different experiments and is not correlated to the composition of the aggregate when the initial geometry ($p_\textrm{ini}$, $q_\textrm{ini}$) is kept constant (see SI). The remarkable result that the work is nearly independent of the composition of the aggregate, is an indication that the number of bonds broken must be nearly constant.

\begin{figure}[]
\begin{center}
\includegraphics[width=0.9\columnwidth,keepaspectratio=true]{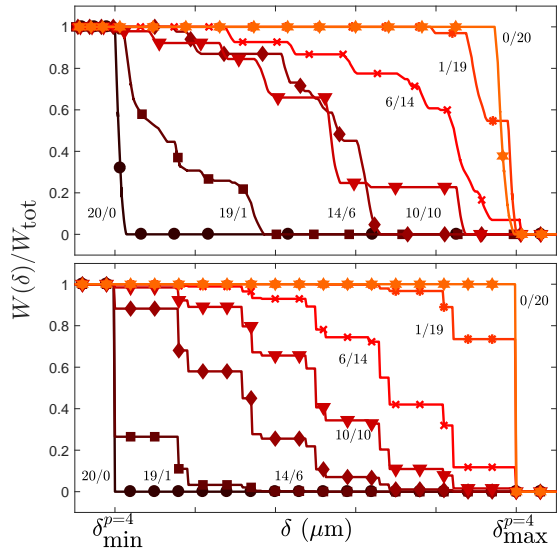}
\caption{\label{fig5} Normalized partial work (see definition in text) as a function of distance, for the $p=4\rightarrow p=3$ transition, for aggregates of different relative compositions (number of small droplets / number of large droplets) as indicated. (top) Experimental results corresponding to the force curves shown in Fig.~\ref{fig2}(a). (bottom) Corresponding theoretical results, according to Eq.~(\ref{Work}). }
\end{center}
\end{figure}

While the total work may be nearly constant, there is an important distinction between the disordered and crystalline systems in how that work is distributed during a $p\rightarrow(p-1)$ transition of the aggregate. To access that information, we consider the partial work $W(\delta)=\int_{\delta}^{\delta_{\textrm{max}}^p}\textrm{d}\delta'\,F_{\textrm{s}}(\delta')$, with $\delta \in [\delta_{\textrm{min}}^p,\delta_{\textrm{max}}^p]$. For the crystals, the bonds are broken simultaneously as the system is driven out of a deep minimum in the landscape. For instance, the crystal made of large droplets breaks near $\delta = \delta_{\textrm{max}}^{p}$ [Fig.~\ref{fig2}(a)], with the normalized partial work going abruptly from zero to one upon compression (\emph{i.e.} decreasing  $\delta$) near that point. This fracture event is detailed in the top panel of Fig.~\ref{fig5} where we plot the normalized work as a function of the inter-pipette distance, $\delta$. The crystal made of small droplets exhibits a similar sudden transition, except that the fracture event happens at $\delta = \delta_{\textrm{min}}^{p}$. In contrast, when defects are introduced, several intermediate steps are observed. For an aggregate with a single defect, a major step (corresponding to the crystalline fraction) is still observed but rapidly fades away as more defects are added. The curves for 6 defects ($\phi=0.3$) and for the model glass ($\phi=0.5$) both show many discrete jumps in the work -- thus, the failure of disordered systems is more progressive and has a much lower yield threshold than for crystals.  
Finally, one can compare the experiment to the theoretical model developed above. In the model, the average normalized partial work is given by the fraction of columns that have a height $H$ larger than $\delta$. Invoking the probability distribution $P$ of column heights, one gets on average: 
\begin{equation}
\frac{W(\delta)}{W_\textrm{tot}}= \int_{\frac{\delta}{\Lambda}}^{\frac{\delta_{\textrm{max}}^p}{\Lambda}}\textrm{d}h\, P(h,\phi,p)\ ,
\label{Work}
\end{equation}
where $\Lambda = \mu(p,\phi = 0.5) = (H_r+H_R)/2$. This expression is plotted in the bottom panel of Fig.~\ref{fig5} for various  compositions, and is consistent with the experimental data. The theory predicts more steps than the experiment, this is because the experiment probes one configuration, while the theory is an average over all the configurations.   We have thus shown that model 2D crystals and glasses are markedly different under compression:  crystals deform elastically until a catastrophic global fracture event occurs, whereas glasses rearrange locally with many intermediate fracture events that each have lower individual yield thresholds. This deviation from the well-established response of a crystal to an external stress has also been observed in a recent analytical study~\cite{Biroli2016}, and it was shown numerically that materials go from brittle to ductile when transitioning from crystal to glass~\cite{Babu2016} -- a fact that is tested here directly with the idealised microscopic experiments. 

\section*{Conclusions}
In summary, by systematically adding disorder in finite-size 2D colloidal crystals, we have studied the crystal-to-glass transition. Upon addition of defects the mechanical properties of the aggregates rapidly transition from crystalline to glassy. The number of force peaks, corresponding to fracture events, increases steeply with the defect fraction, before saturating to the glass value. Additionally, the yield energy as a function of disorder has been investigated. We find that for a 2D crystal, a high energy barrier must be overcome, while glasses fracture progressively through failure in many small steps. In the system studied the adhesion energy between particles exceeds the thermal energy, thus the aggregates correspond to a glass or a crystal well below the solid-melt transition temperature. The fracture events observed reflect the substructure introduced by  disorder in the underlying energy landscape. This is consistent with the brittle failure of crystals as opposed to the plasticity of glasses. A minimal analytical model captures the essential experimental features. From the combination of experiments and theory, we quantify the crystal-to-glass transition using macroscopic yield observables that are consistent with a simple microscopic picture.

\section*{Materials and Methods}

\subsection*{Experimental setup} A chamber ($55 \times 30$~mm) is made of two glass slides separated by a gap of 2.5~mm, which is $10^3$ times greater than the size of droplets. The chamber is filled with an aqueous solution of sodium dodecyl sulfate (SDS) at 3\% and NaCl at 1.5\%. This concentration of SDS leads to the formation of micelles acting as a depletant resulting in a short-ranged attraction between the droplets~\cite{Bibette1992}. The chamber is placed atop an inverted optical microscope for imaging while the aggregates are compressed.  

\subsection*{Micropipettes} Three small micropipettes are inserted into the chamber: the ``droplet pipette'', ``pushing pipette", and ``force-sensing pipette". Pipettes were pulled from glass capillaries (World Precision Instruments, USA) with a pipette puller (Narishige, Japan) to a diameter of about 10~$\mu$m over several centimeters in length. The ``droplet pipette'' produces monodisperse droplets, with size directly proportional to the tip radius of the pipette, using the snap-off instability~\cite{Barkley2016}. The droplets are buoyant and form a 2D aggregate under the top glass slide [Fig.~1(b)]. The ``pushing pipette" is short and stiff and is used to compress the aggregate. The pushing pipette is affixed to a translation stage and its speed set to 0.3~$\mu$m/s for all experiments. The ``force-sensing pipette" is a long compliant pipette, and its deflection is used to measure forces applied to the aggregate~\cite{Backholm2013, Backholm2019}. To be sensitive to forces as small as $\approx 100$~pN, the force-sensing pipette needs to be long ($\approx 3$~cm) and thin ($\approx 10$~$\mu$m). This long straight pipette is locally and temporarily heated to soften the glass such that it can be shaped to fit  within the small chamber [see pipette (iii) in Fig.~\ref{fig1}~(a)].

\subsection*{Preparation of the aggregates} Aggregates of oil droplets are assembled droplet-by-droplet and thus can be prepared into any arbitrary shape.

\subsection*{Cross-correlation analysis} The distance between  the pushing pipette and the force-sensing pipette, $\delta$, is measured using cross-correlation analysis between images with a precision of $\sim 0.1$ $\mu$m~\cite{Backholm2013, Backholm2019}. Additionally, correlation analysis provides the deflection of the force-sensing pipette, which is converted to a force using the calibrated spring constant $k_{\textrm{p}} =1.3 \pm 0.1$ nN/$\mu$m of the pipette~\cite{Backholm2013, Backholm2019}. The typical uncertainty on the force is: $\delta F/F \approx 2\%$.

\section*{Acknowledgements} 
The authors thank Maxence Arutkin, Matilda Backholm, and James Forrest for valuable discussions, as well as Yilong Han for sharing a preliminary draft on a similar topic. Financial support from NSERC (Canada), the Joliot chair from ESPCI Paris, and the Global Station for Soft Matter, a project of Global Institution for Collaborative Research and Education at Hokkaido University is gratefully acknowledged. The work of ERW was supported by the National Science Foundation (CBET-1804186).

% Bibliography

%*******************

\newpage
\onecolumngrid
\appendix
\section*{Supplementary Information}
\renewcommand{\theequation}{S\arabic{equation}}

\section{Height of a column and probability}
In this section, the theoretical model used to predict the number of peaks in the force measurement as a function of $\phi$ is derived. In this calculation, each transition from $p$ rows of droplets to $(p-1)$ rows is studied individually. In the following, $p$ and $q$ are constant values.
 
The theoretical model developed for this study is based on geometrical arguments. An assembly of droplets is compressed if its lateral unstrained extent is larger than the spacing between the pipettes. The aggregate is modeled as $q$ independent columns of height $H_i$ stacked next to each other, the index $i$, going from $1$ to $q$, is labelling the columns. The total height of a column depends the composition of droplets. For a crystal, all the columns are the same so they break at the same time, which results in a single peak in the force measurement. When defects are introduced, large droplets are substituted by small ones (or {\it vice versa}). Columns constituting the aggregate now have different heights and break for different values of $\delta$ resulting in several peaks in the force measurement. 

\begin{figure}[h]
\centering
\includegraphics[width=5cm]{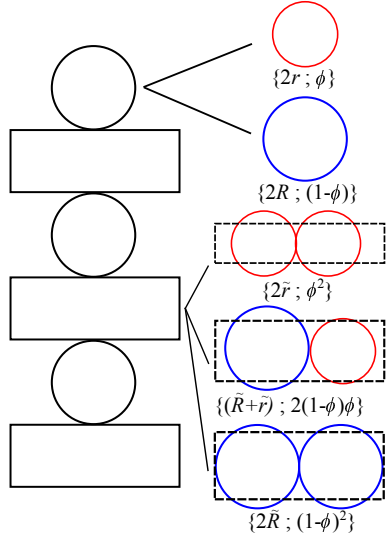}
\caption{Schematics of the columns with $p=6$ considered in the theoretical model. The left part shows how rectangles and circles are assembled to build a column. The right part shows the different choices for circles and rectangles along with their probability to appear.}
\label{column} 
\end{figure}

A column is made of alternating layers of two droplets, which are modelled as a \emph{rectangle}, and single droplets, modelled as \emph{circles}, as shown in Fig.~\ref{column}. The number fraction $\phi$ of small droplets in an assembly of $N_\textrm{small}$ small droplets and $N_\textrm{tot}-N_\textrm{small}$ large droplets is defined as: $\phi = N_\textrm{small}/N_\textrm{tot}$. Depending on the composition of the two droplets,  the rectangles can take three heights \{2$\tilde{R}$, 2$\tilde{r}$, ($\tilde{R}$+$\tilde{r}$)\} with probabilities \{$(1-\phi)^2$, $\phi^2$, $2(1-\phi)\phi$\} respectively. The circles can only have two diameters resulting in two distinct heights  \{$2R$, $2r$\} with probabilities \{$(1-\phi)$, $\phi$\} respectively. Finally, we take the relation between ($R$, $r$) and ($\tilde{R}$, $\tilde{r}$) to be a geometrical factor $\alpha$. It is the sum of the heights of the rectangles and the heights of the circles that determine the overall height of a column as shown in Fig.~\ref{column}.

Random walk statistics can be applied to this model. To simplify, we consider $p$ being even. The results for $p$ being odd would be similar. Thus, for the even case, building such a column is equivalent to two random walks of p/2 steps: one with the circles and one with the rectangles. Using the random walk statistics formalism \cite{Sokolov2011}, we can express the probability  $P_\textrm{circ}(H_1,\phi,p,r,R)$ of finding a height $H_1$ by stacking $p/2$ circles of two different sizes for a given $\phi$, and $P_\textrm{rect}(H_2,\phi,p,\tilde{r},\tilde{R})$, the probability of finding a height $H_2$ by stacking $p/2$ rectangles of three different heights:

\begin{equation}
\left\{
  \begin{array}{ll}
        P_\textrm{circ}(H_1,\phi,p,r,R)=\frac{1}{2\pi}\int_{-\pi}^{\pi}(\phi e^{2i\theta r}+(1-\phi)e^{2i\theta R})^{\frac{p}{2}}e^{-i\theta H_1}\textrm{d}\theta\\
       P_\textrm{rect}(H_2,\phi,p,\tilde{r},\tilde{R})=\frac{1}{2\pi}\int_{-\pi}^{\pi}(\underbrace{\phi^2 e^{2i\theta \tilde{r}}+(1-\phi)^2e^{2i\theta \tilde{R}}+2\phi(1-\phi)e^{i\theta (\tilde{R}+\tilde{r})}}_{(\phi e^{i\theta \tilde{r}}+(1-\phi)e^{i\theta \tilde{R}})^2})^{\frac{p}{2}}e^{-i\theta H_2}\textrm{d}\theta
    \end{array}
\right.
\label{RW}
\end{equation}

\begin{equation}
\left\{
    \begin{array}{ll}
        P_\textrm{circ}(H_1,\phi,p,r,R)=\sum_{k=0}^{\frac{p}{2}}{{\frac{p}{2}}\choose{k}}\phi^k(1-\phi)^{\frac{p}{2}-k}\delta(2kr+2(\frac{p}{2}-k)R-H_1)\\
        \\
        P_\textrm{rect}(H_2,\phi,p,\tilde{r},\tilde{R})=\sum_{l=0}^{p}{{p}\choose{l}}\phi^l(1-\phi)^{p-l}\delta(l\tilde{r}+(p-l)\tilde{R}-H_2).
    \end{array}
\right.
\label{Proba piles sep}
\end{equation}

It turns out that the random walk of $p/2$ steps with three different step sizes is equivalent to $p$ steps of two different sizes (see Eq.~\ref{RW}). The Dirac $\delta$ function in Eq.~\ref{Proba piles sep} is a geometrical constraint on the total height. Only the combinations of droplets that leads to the right total heights $H_1$ and $H_2$ are considered. The distribution of probability of the total height $H$ is the convolution product of $P_\textrm{circ}(H_1,p,\phi,r,R)$ and $P_\textrm{rect}(H_2,p,\phi,\tilde{r},\tilde{R})$:

\begin{equation}
P(H,\phi,p,r,R,\tilde{r},\tilde{R})=\int_0^{\infty}P_\textrm{circ}(H-H_2,\phi,p,r,R)P_\textrm{rect}(H_2,\phi,p,\tilde{r},\tilde{R})\,\textrm{d}H_2.
\end{equation}
Using Eq. \ref{Proba piles sep} we find:
\begin{equation}
P(H,\phi,p,r,R,\tilde{r},\tilde{R})=\sum_{k=0}^{\frac{p}{2}}\sum_{l=0}^{p}{{\frac{p}{2}}\choose{k}}{{p}\choose{l}}\phi^{k+l}(1-\phi)^{\frac{3p}{2}-k-l}\delta(l\tilde{r}+(p-l)\tilde{R}+2kr+2(\frac{p}{2}-k)R-H).
\label{proba hauteur}
\end{equation}
To simplify the notation we consider $r$, $R$, $\tilde{r}$ and $\tilde{R}$ fixed so the probability distribution is $P(H,\phi,p)$.  Eq.~\ref{proba hauteur} is used to calculate numerically the discrete distribution presented in the main text (Fig.~3(c)-(e) histograms).  The height $H$ can take discrete values $H_i$, with probability $P_i=P(H_i,\phi,p)$, ranging from $H_r$ for a column made of small droplets ($\phi=0$) to $H_R$ for a column made of large droplets ($\phi=1$). The total number of different heights $H_i$ only depends on $p$ and is noted $m_p$. Finally, the height $H$ is renormalized as follows $h=2H/(H_r+H_R)$. With this renormalization, a column made of 50\% large droplets and 50\% small droplets ($\phi = 0.5$) has a dimensionless height $h = 1$.

\section{Number of peaks}

The column model gives access to the probability $P_i$ of finding the height $H_i$ in an aggregate for any fraction of defects $\phi$. The number of peaks in the force measurement is calculated from the height distribution. Let us denote the average number of force peaks during the compression of an aggregate with $p$ rows to an aggregate with $(p-1)$ rows by $N^{p}(\phi,q)$. Observing a single peak in the force measurement means that all the columns share the same height. Measuring two peaks means that there are two and only two different heights. Thus $N^{p}(\phi,q)$ corresponds to the average number of \emph{different} heights composing an aggregate of $p$ rows and $q$ columns. 

For a given fraction of defects $\phi$, an aggregate of size $p\times q$ is built by choosing randomly $q$ columns from a pool of columns. Correlation between two adjacent columns are neglected. Experimentally, a small fraction of the peaks in the force measurement is due to the correlation between columns but most of the peaks are indeed due to compression of independent columns. From the discrete probability distribution, Eq.~\ref{proba hauteur}, there is a finite number $m_p$ of possible heights $H_i$ with a non-zero probability. To predict the number of peaks we calculate the probability $A_{n}(\phi,p,q)$ of finding strictly $n$ different columns heights in an aggregate of size $p\times q$ at a given fraction of defects $\phi$.

\begin{figure}[h]
\centering
\includegraphics[width=12cm]{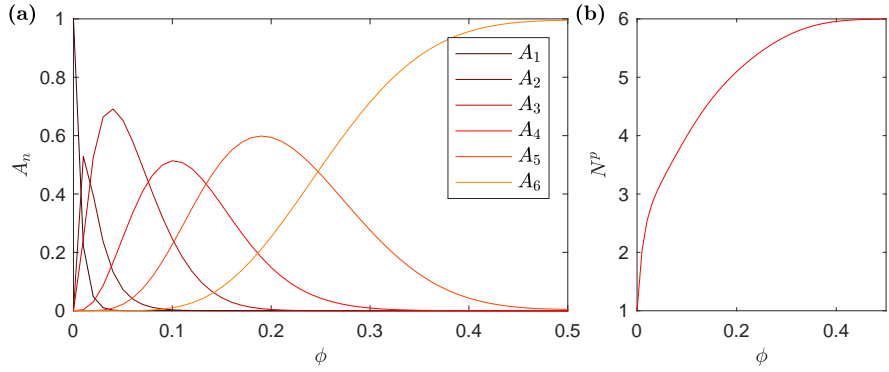}
\caption{(a) Probability distribution $A_{n}$ as a function of $\phi$ for an aggregate made of $p=2$ rows and $q=50$ droplets per row. (b) Prediction of the average number of peaks in the force measurement as a function of $\phi$ based on the distribution $A_{n}$. Note that we only plot the function for $\phi \in [0,0.5]$ by symmetry about $\phi=0.5$.
} 
\label{An peaks}
\end{figure}

Building an aggregate is equivalent to drawing $q$ columns which can take $m_p$ different heights $H_i$ with probability $P_i$. $\{ijk...\}_{n}$ defines $n$ different numbers between $1$ and $m_p$. Let $\tilde{P}_{\{ijk...\}_{n}}$ denote the probability that the aggregate is composed only of the $n$ heights $\{H_i,H_j,H_k,...\}$ and each height appears at least once. As the order in which the heights are drawn is not important, one gets:
\begin{equation}
A_{n}(\phi,p,q)=\frac{\Theta(q-n)}{n!}\sum_{\{ijk...\}_n\subset\llbracket 1,m_p\rrbracket}\tilde{P}_{\{ijk...\}_{n}};
\label{eq:An_def}
\end{equation}
 where $\Theta$ is the Heaviside function and $\sum_{\{ijk...\}_n\subset\llbracket 1,m_p\rrbracket}$ denotes the sum over all the $n$-tuples $\{ijk...\}_n$ in $\llbracket 1,m_p\rrbracket$.
We define $P_{\{ijk...\}_{n}}$ as the probability to draw one of the $n$ heights $\{H_i,H_j,H_k,...\}$, and obtain:
\begin {equation}
P_{\{ijk...\}_{n}}=\sum_{\kappa\in\{ijk...\}}P_\kappa.
\end{equation}
Hence, the probability that an aggregate of $q$ columns is composed only of the heights $\{H_i,H_j,H_k,...\}$ is given by $(P_{\{ijk...\}_{n}})^{q}$. However, this probability is not equal to $\tilde{P}_{\{ijk...\}_{n}}$ since it does not take into account that each height must appear at least once. The difference between $(P_{\{ijk...\}_{n}})^q$ and $\tilde{P}_{\{ijk...\}_{n}}$ is the probability that one or more of the heights  $\{H_i,H_j,H_k,...\}$ does not appear. To calculate $\tilde{P}_{\{ijk...\}_{n}}$ we subtract from $(P_{\{ijk...\}_{n}})^q$ the probabilities that the aggregate is only composed of $n-\kappa$ different types of columns of heights $\{H_a,H_b,H_c,...\}$ with $\{abc...\}\subset\{ijk...\}_{n}$ summed over all the possible $(n-\kappa)$-tuples in $\{ijk...\}_n$ and summed over all the $\kappa$ from $1$ to $n-1$:
\begin{equation}
\tilde{P}_{\{ijk...\}_{n}}=(P_{\{ijk...\}_{n}})^q-\sum_{\kappa=1}^{n-1}~\sum_{\{abc...\}_{n-\kappa}\subset\{ijk...\}_n}~\tilde{P}_{\{abc...\}_{n-\kappa}}.
\end{equation}
Noticing that the sum over the $\kappa$ and the other sums can be switched and using Eq.~\ref{eq:An_def}, one finds: 
\begin{equation}
A_{n}(\phi,p,q)=\frac{\Theta(q-n)}{n!}\sum_{\{ijk...\}_n\subset\llbracket 1,m_p\rrbracket}(P_{\{ijk...\}_{n}})^q-\frac{1}{n!}\sum_{\kappa=1}^{n-1}~\underbrace{\sum_{\{ijk...\}_n\subset\llbracket 1,m_p\rrbracket}~\sum_{\{abc...\}_{n-\kappa}\subset\{ijk...\}_n}~\tilde{P}_{\{abc...\}_{n-\kappa}}}_{B_\kappa}.
\label{bk}
\end{equation}
Let us focus on the second term of the right hand side, called $B_\kappa$, in Eq.~\ref{bk}. $\tilde{P}_{\{abc...\}_{n-\kappa}}$ depends on $n-\kappa$ indices and it is summed over $n$ indices. So if we fix the $n-\kappa$ indices that $\tilde{P}_{\{abc...\}_{n-\kappa}}$ depends on, it will appears ${{n}\choose{\kappa}}$ times. Thus:
\begin{equation}
B_\kappa=\sum_{\{ijk...\}_{n}\subset\llbracket 1,m_p\rrbracket}{{n}\choose{\kappa}}\tilde{P}_{\{ijk...\}_{n-\kappa}}.
\end{equation}
By splitting the sum into two parts, one finds:
\begin{equation}
B_\kappa=\kappa!{{n}\choose{\kappa}}\bigg[\prod_{\beta=\kappa}^{n-1}(m_p-\beta)\bigg]A_{\kappa},
\end{equation}
leading to:
\begin{equation}
A_{n}(\phi,p,q)=\frac{\Theta(q-n)}{n!}\sum_{\{ijk...\}_n\subset\llbracket 1,m_p\rrbracket}(P_{\{ijk...\}_{n}})^{q}-\frac{1}{n!}\sum_{\kappa=1}^{n-1}{{n}\choose{\kappa}}\kappa !\bigg[\prod_{\beta=\kappa}^{n-1}(m_p-\beta)\bigg]A_{\kappa}.
\label{A_n}
\end{equation}
The average number of peaks $N^{p}(\phi,q)$ is given by:
\begin{equation}
N^{p}(\phi,q)= \sum_{n=1}^{m_p} n A_{n}(\phi,p,q).
\label{N discrete}
\end{equation}
Equation~\ref{A_n} and \ref{N discrete} can be evaluated numerically, see Fig.~\ref{An peaks}.
The importance of this distribution $A_{n}(\phi,p,q)$ can be easily understood for both extreme values of $\phi$. If $\phi = 0$, there is only one possible height for the column meaning that $A_{n}(\phi=0,p,q)=\delta_{1n}$, where $\delta_{ij}$ is the Kronecker symbol. On the other hand, if $\phi =0.5$, it is unlikely to find only one height so $A_1 \simeq 0$. It is more likely to find all the different heights in the aggregate leading to $A_{m_p} \simeq 1$. This is illustrated in Fig.~\ref{An peaks}(a) which shows the probabilities, $A_{n}$, for $p=2$ (with $n$ takes values from 1 to $m_2=6$) as a function of $\phi$. Note that in Figs.~\ref{An peaks} (a)-(b) we restrict the range to $\phi \in [0,0.5]$ as the function is symmetric about $\phi=0.5$. For $\phi=0$, only $A_1 \ne 0$. As $\phi$ increases, finding two different heights becomes more likely and $A_2$ becomes dominent. For $\phi=0.5$, it is very likely to find the maximum number of columns, $m_2=6$, in the aggregate and $A_6 \simeq 1$. 

\subsection{Finite size effect}

The number of force peaks is a function of the size of the cluster: in the model, the number of force peaks depends explicitly on the number of columns $q$ since $A_n$ depends on $q$ (in Fig.~\ref{An peaks}, $q=50$).
In an infinitely large cluster ($q\rightarrow\infty$) all the heights will appear and so : $$N^{p}(\phi> 0,q\rightarrow\infty)=m_p,$$ with $m_p$ the number of possible heights one can get with the random packing described previously at a given $p$. For a finite $q$, one has $N^{p}(\phi,q)<m_p$ because all the possible heights will not appear simultaneously in the same cluster. As a simple consequence, one has: $N^{p}(\phi,q_1)<N^{p}(\phi,q_2),\;\textrm{for } q_1<q_2 .$

\begin{figure}[h]
\centering
\includegraphics[width=12cm]{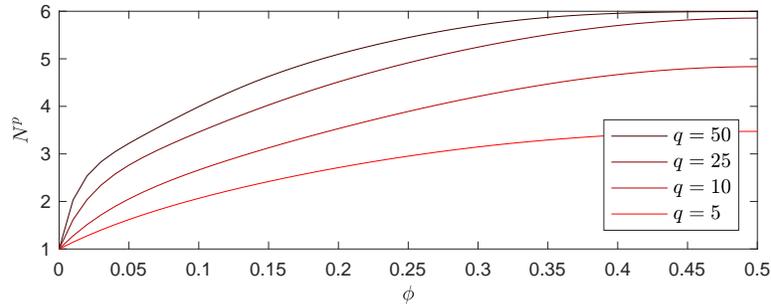}
\caption{Impact of the size of the aggregate on the number of peaks in the force measurement for $p=2$ and different values of $q$.}
\label{sizeeffect}
\end{figure}

Moreover, even if the total number of different possible heights $m_p$ is large, we cannot find more different heights than the number of columns, $q$. This is the reason for the Heaviside function in the definition of $A_n$.
For the experimental aggregates, $q$ varies from 3 to 15. In particular for $p=3$ or $p=4$, the number of columns is usually $\sim$ 5 and the value of $q$ gives an upper limit for the number of force peaks. 
Fig.~\ref{sizeeffect} shows the impact of the number of columns on the number of peaks for $p=2$. 

\subsection{A simpler approach to estimate the number of force peaks}

The approach discussed above predicts accurately the average number of force peaks observed during the compression of an aggregate. However the number of force peaks estimated for a transition $p\rightarrow p-1$ is strongly dependent of the size of the cluster. In this section, we take a simpler approach that leads to an analytical expression for the number of force peaks observed during the compression of a cluster. In addition, with this approach we are able to define a quantity that allows us to renormalize our results with respect to the size of the cluster. This analytical expression characterizes the transition of a cluster from being crystal-like to glass-like.

We define the excess number of force peaks as the number of force peaks for a given $\phi$ compared to the number of force peaks observed in the crystal case $N^{p}(\phi=0,q)=1$  for the same transition $p\rightarrow p-1$: $N^{p}(\phi,q) - 1$. This quantity can be normalized by its value for a glassy case where $\phi=0.5$: $N^{p}(\phi = 0.5,q) - 1$.
The normalized quantity quantifies how crystalline or glassy a cluster is, and takes values ranging from $0$ for a crystal to $1$ for a glass.

\begin{figure}[h]
\centering
\includegraphics[width=12cm]{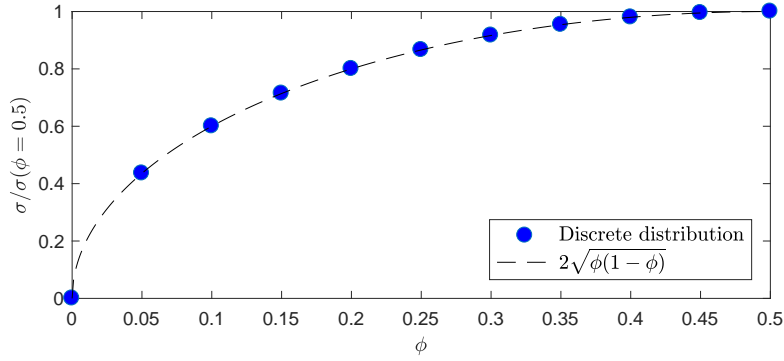}
\caption{Comparison between the analytical expression for the standard deviation of the height distribution and the numerically calculated values for different $\phi$ for $p=2$.}
\label{disc-cont}
\end{figure}

Instead of numerically calculating the average number of different column heights in a $p\times q$ cluster, we propose the following statistical argument: the average number of different heights in a $p\times q$ cluster is well approximated by the number of different \textit{highly probable heights} in the probability distribution of heights. To define if a height $H_i$ is highly probable, one has to invoke a threshold for the probability, $P(H_i)$, which is strongly dependent on the total number of columns, $q$, in the cluster. The larger $q$ is, the smaller the threshold must be, and should go to zero in the limit of infinitely large clusters ($q\rightarrow\infty$). 
Since the columns are built as a 1D random walk, the distribution follows a binomial law and the number of probable heights can be characterized using the standard deviation $\sigma$ of the height distribution. Using $\sigma$ to define the threshold, the excess number of peaks is then directly proportional to the width of the height distribution. As we are interested in the ratio between the excess number of peaks at a given fraction of defects, $\phi$, and its maximum value, observed at $\phi = 0.5$, the choice of the threshold is not critical, for the range of $q$ values explored in the experiments.  This leads to:
\begin{equation}
\frac{N^{p}(\phi,q) - 1}{N^{p}(\phi = 0.5,q) - 1} \simeq \frac{\sigma(\phi,p)}{\sigma(\phi=0.5,p)}.
\label{Np}
\end{equation}
The standard deviation, $\sigma$, as well as the average value, $\mu$, of the continuous height distribution can be calculated analytically in this simplified approach. We find $\mu(\phi,p)=p(\alpha+1)[(1-\phi)R+r\phi]$ and $\sigma^2(\phi,p)=p(2+\alpha^2)(R-r)^2\phi(1-\phi)$. The analytical expression for $\sigma$ as a function of $\phi$ is tested against the numerically calculated values from the discrete model, described in the previous sections, for different values of $\phi$. Figure \ref{disc-cont} shows perfect agreement between both approaches for $p=2$.

\begin{figure}[h]
\includegraphics[width=12cm]{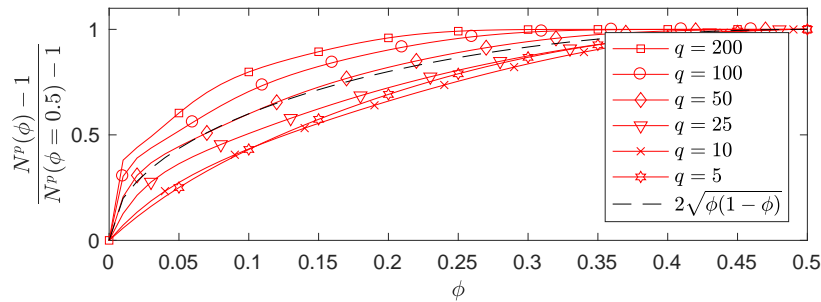}
\caption{Comparison between the average number of peaks predicted by the discrete calculation (for different values of $q$) and the continuous approximation, for $p=2$. The continuous model does not take into account the number of droplets per row $q$ in the aggregate.}
\label{comparisonfig}
\end{figure}

Using the expression found for the standard deviation, $\sigma$, Eq.~\ref{Np} can be rewritten as:
\begin{equation}
\frac{N^{p}(\phi,q) - 1}{N^{p}(\phi = 0.5,q) - 1} \simeq 2\sqrt{\phi(1-\phi)}.
\label{comparison}
\end{equation}
Note that the result is now independent of the size of the cluster. Indeed, this ratio simply compares the excess number of peaks to its maximum value, but it does not predict the exact number of peaks observed in the force curves.
This simplified approach can be tested against the discrete model by comparing the number of peaks predicted by each model with $p=2$. The left hand side of Eq.~\ref{comparison} is calculated numerically for the discrete model and compared to $2\sqrt{\phi(1-\phi)}$ as shown in Fig.~\ref{comparisonfig}. Both models are in good agreement for the number of peaks as long as $q<50$. The analytical prediction overestimates the number of force peaks in the range of $q$ values experimentally explored.

\section{Total number of peaks $N^{p_{\textrm{ini}}\rightarrow 1}_{q_{\textrm{ini}}}(\phi)$}

In the main text, we compare the experimental results and the total number of peaks when compressing a cluster initially made of $p_\textrm{ini}$ rows and  $q_\textrm{ini}$ columns to a single row ($p=1$), at a given percentage in defects $\phi$: $N^{p_{\textrm{ini}}\rightarrow 1}_{q_{\textrm{ini}}}(\phi)$. Equation 2 in the main text is obtained by summing Eq.~\ref{comparison} over the different transitions:
\begin{equation}
N^{p_{\textrm{ini}}\rightarrow 1}_{q_{\textrm{ini}}}(\phi)= \sum_{p=2}^{p_\textrm{ini}} N^{p}(\phi,q) = 2\sqrt{(1-\phi)\phi} \sum_{p=2}^{p_\textrm{ini}} (N^{p}(\phi=0.5,q) - 1)  + \sum_{p=2}^{p_\textrm{ini}} 1.
\label{15}
\end{equation}
During these transitions, the total number of droplets $N_\textrm{tot}$ is conserved and thus $p\times q = p_\textrm{ini}\times q_\textrm{ini}$. Noticing that $\sum_{p=2}^{p_\textrm{ini}} 1$ is the number of peaks observed when compressing a crystal initially made of $p_\textrm{ini}$ rows, this quantity is independent of $q_{\textrm{ini}}$ and will be noted $N^{p_{\textrm{ini}}\rightarrow 1}(\phi=0)$. Equation \ref{15} can then be written as:
\begin{equation}
N^{p_{\textrm{ini}}\rightarrow 1}_{q_{\textrm{ini}}}(\phi)= 2\sqrt{(1-\phi)\phi} (N^{p_{\textrm{ini}}\rightarrow 1}_{q_{\textrm{ini}}}(\phi=0.5)- N^{p_{\textrm{ini}}\rightarrow 1}(0))  + N^{p_{\textrm{ini}}\rightarrow 1}(0),
\end{equation}
leading to:
\begin{equation}
\frac{N^{p_{\textrm{ini}}\rightarrow 1}_{q_{\textrm{ini}}}(\phi) -  N^{p_{\textrm{ini}}\rightarrow 1}(0)}{ N^{p_{\textrm{ini}}\rightarrow 1}_{q_{\textrm{ini}}}(\phi=0.5)- N^{p_{\textrm{ini}}\rightarrow 1}(0)}= 2\sqrt{(1-\phi)\phi}.
\label{renorm}
\end{equation}

Finally, we define $\Delta N(\phi,p_\textrm{ini},q_\textrm{ini}) = N^{p_{\textrm{ini}}\rightarrow 1}_{q_{\textrm{ini}}}(\phi)-  N^{p_{\textrm{ini}}\rightarrow 1}(0)$ as the average excess number of peaks observed when compressing an aggregate with a defect fraction $\phi$ in comparison to a crystal of same geometry $p_\textrm{ini}\times q_\textrm{ini}$. The maximum excess number of peaks $\Delta N_\textrm{max} = N^{p_{\textrm{ini}}\rightarrow 1}_{q_{\textrm{ini}}}(\phi=0.5) - N^{p_{\textrm{ini}}\rightarrow 1}(0)$ corresponds to the excess number of peaks observed when compressing the most disordered aggregates (model for a glass, $\phi = 0.5$). The ratio of these two quantities $\Delta N/\Delta N_\textrm{max}$ does not depend on the size of the cluster $p_\textrm{ini}\times q_\textrm{ini}$ but only on the fraction of defects $\phi$. We can thus write  Eq.~\ref{renorm} in a simpler form and obtain Eq.~2 of the main text:
\begin{equation}
\frac{\Delta N}{\Delta N_\textrm{max}} (\phi)= 2\sqrt{(1-\phi)\phi}.
\label{eq2}
\end{equation}
The maximum number of peaks, $N_\textrm{max}^{p_\textrm{ini}\rightarrow 1}(q) = N^{p_\textrm{ini}\rightarrow 1} (\phi=0.5,q)$, depends on the system size as it has been shown in the previous section. The experimental value is obtained by fitting Eq.~\ref{eq2} to each set of data presented in Fig. 3(a).

\section{Size of the droplets in the crystal-to-glass transition study}

The data shown in Figs.~4(a)-(b), in the main text, come from four different sets of experiments. The droplets used during a given set of experiments are the same, while the defect fraction is varied. New droplets were produced for each new set. Table~\ref{dropletsize} summarizes the sizes of the droplets used in these experiments. 

\begin{table}[h]
\begin{center}
\caption{Size of the droplets used for the crystal-to-glass transition study - Fig.~4 in the main text.}
\begin{tabular}{c|c|c|c}

Data set & Points colour & $R$ ($\mu$m) & $r$ ($\mu$m)  \\
\hline
1 & red & $21.5 \pm 0.2$ & $19.1\pm 0.2$ \\

2 & salmon & $21.4 \pm 0.4$ & $19.1\pm 0.4$ \\

 3 & light blue & $20.9 \pm 0.3$ & $17.3\pm 0.5$ \\

 4 & dark blue & $25.1 \pm 0.3$ & $19.2\pm 0.3$ \\

\end{tabular}

\label{dropletsize}
\end{center}
\end{table}

\section{Work analysis}
In the main text, we study how the work is distributed along a compression as a function of the composition of the aggregate. This analysis relies on the assumption that the total work for a given transition does not depend on $\phi$. We found that within the uncertainty of the experiments, the total work for the transition $p \rightarrow (p-1)$ is constant and is not correlated to the fraction of defects $\phi$. Table~\ref{work} summarizes the total work, $W_\textrm{tot}$, exerted to go from four to three rows for the different aggregates. 

\begin{table}[h]
\begin{center}
\caption{Total work needed to transition from $p=4$ to $p=3$ for different compositions.}
\begin{tabular}{c|c|c}
Composition & $\phi$ & $W_\textrm{tot}$ (fJ) \\
\hline
  20/0 & 0 & 1.2\\
  19/1 & 0.05 & 2.8\\
  14/6 & 0.3 & 2.8\\
  10/10 & 0.5 & 2.1\\
  6/14 & 0.3 & 3.1\\
  1/19 & 0.05 & 1.5\\
  0/20 & 0 & 2.0\\
\end{tabular}
\label{work}
\end{center}
\end{table}

\end{document}